\documentclass{kluwer}    
\usepackage[]{graphicx}

\newdisplay{guess}{Conjecture}

\begin{document}                                                                                   
\begin{article}
\begin{opening}
\title{The evolution of cosmic star formation, metals and gas}
\author{Francesco \surname{Calura} and Francesca \surname{Matteucci} }
\runningauthor{F. Calura and F. Matteucci}
\institute{Dipartimento di Astronomia, Universit\'{a} di Trieste, via
	G. B. Tiepolo 11, 34131 Trieste, Italy}

\begin{abstract}
We reconstruct the history of the cosmic star formation
as well as the cosmic production of metals
in the universe by means of
detailed chemical evolution models for galaxies of different morphological 
types. 
We consider a picture of coeval, non-interacting evolving
galaxies where ellipticals experience intense
and rapid starbursts within the first Gyr after their formation, and spirals 
and irregulars continue to form stars at lower rates up to the present time.
We show that spirals are the main contributors to the decline of the luminosity
density in all bands between $z=1$ and $z=0$.

\end{abstract}

\end{opening}           

\section{The evolution of the luminosity density}  

The study of the galaxy luminosity density in various bands is fundamental in understanding how galaxies formed 
and evolved in the universe. 
We start from the B-band luminosity function (LF) in the local universe by Marzke et al. (1998)
and by means of chemo-photometric models of galaxy evolution we aim at reconstructing the history of the luminous
matter in the universe. Our models calculate the chemical evolution of galaxies of different morphological type, i.e.
ellipticals, spirals and irregulars. Elliptical galaxies form by means of a rapid collapse of a gas cloud occurring at high redshift,
which gives rise to an intense starburst (Matteucci 1992).
Spirals galaxies form through a double-infall process (Chiappini et al. 1997) and evolve with a continuous star formation history,
whereas irregular galaxies form through continuous accretion of gas and form stars at a slower rate.
For a detailed description of the chemical evolution models employed in our analysis, see Calura et al. (2002).
A photometric model (Jimenez et al. 1998) is matched to our chemical evolution models,
which allows us to calculate self-consistently the photometric evolution of the stellar populations in each galaxy type.
We investigate two different initial mass functions: the Salpeter and the Scalo.
We compute the B-band LF at a given redshift 
by means of the evolutive corrections obtained with our chemo-spectrophotometric code, and  	   
we obtain the LF in other bands (U, I, K) according to the computed galaxy colors. 
We then integrate the LFs assuming a Schechter form.	   
In figures 1 and 2 we compare the results of our calculations with observations at various wavelengths
in the case of galaxy formation at $z_{f}=5$ and $z_{f}=10$, respectively. The adopted cosmology is the 
``concordance'' $\Lambda$CDM, characterized by $\Omega_{m}=0.3$, $\Omega_{\Lambda}=0.7$ with $h=0.65$.
The agreement with the data is good with both the adopted IMFs.
Our main result is that spiral galaxies are the main contributors to the decline of the luminosity density in all bands 
between $z=1$ and $z=0$. 
 
\begin{figure} 
\centerline{\includegraphics[width=17pc]{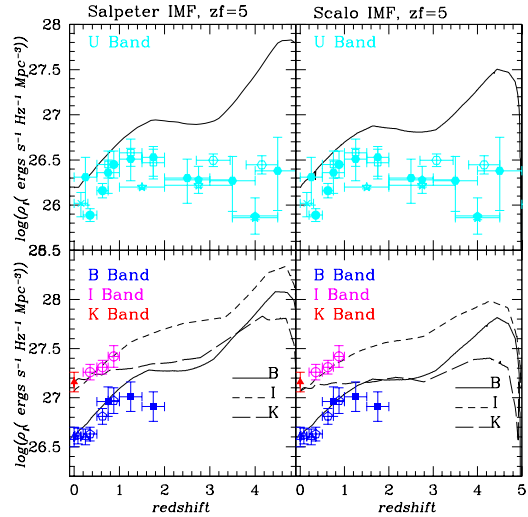}} 
\caption{\tiny Predicted and observed luminosity densities in various bands.
solid exagons from Pascarelle et al. (1999, $1500$ $\mathring{A}$), cross from Treyer et al. (1998, $2000$ $\mathring{A}$),
stars from Massarotti et al. (2002, $1500$ $\mathring{A}$), 
solid penthagons from Madau et al. (1998, $1500$ $\mathring{A}$), solid circles from Lilly et al. (1996, $2800$
$\mathring{A}$), open squares from Connolly et al. (1997, $2800$ $\mathring{A}$).
Open triangles from Ellis et al. (1996,$4400$ $\mathring{A}$), open penthagons from Lilly et al. (1996, 
$4400$ $\mathring{A}$), solid squares from Connolly et al. (1997, $4400$ $\mathring{A}$), solid triangle from Gardner et al.
 (1997, $2.2$ $\mu$), open circles from Lilly et al. (1996, $1$ $\mu$). 
}
\end{figure}

\begin{figure} 
\centerline{\includegraphics[width=17pc]{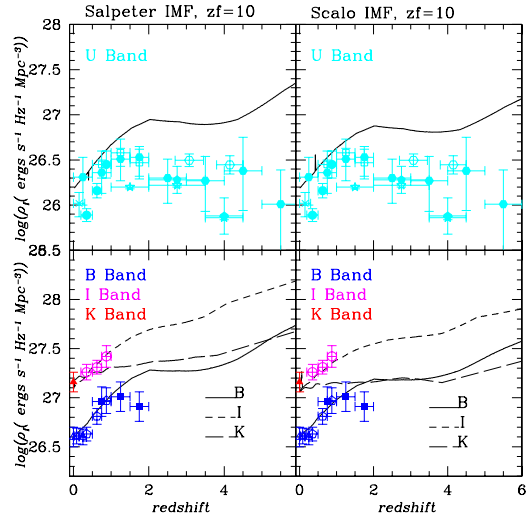}} 
\caption{Data as in the previous figure.}
\end{figure}

\section{The cosmic SFR and HI density} 
Figure 3a shows the evolution of the predicted and measured cosmic SFR density in the case of a Salpeter IMF and galaxy
formation at $z_{f}=5$. 
Our predictions well reproduce the strong evolution between $z=0$ and $z=1$. During this long period
spiral galaxies are the most active forming sites in the universe. The peak at high redshift is due to the intense
starbursts occuring in elliptical galaxies, which at later times stop contributing significantly to the total SFR density.
Irregular galaxies are the least intense star forming sites at all epochs. 
The observations at high redshift do not confirm the existence of the peak predicted by our models, but it is widely
known that the observations at short wavelengths which allow to detect the star formation activity are likely to suffer 
dust-extinction problems (Adelberger \& Steidel 2000, Massarotti et al. 2002).
Thus, if massive star formation occurred in galaxies strongly obscured by dust, it could remain unseen at short wavelengths.
The light could be re-emitted by the dust in the sub-mm band, where observations could be fundamental in 
determining the true existence of such a peak. The total amount of metals produced at early epochs, 
corresponding to the redshift range
where most of the high z abundance measures have been performed,
is given by:
\begin{math} 
\int_{11Gyr}^{13 Gyr} \dot{\rho_{Z}}(t)dt = 8.5 \times 10^{6} M_{\odot} Mpc^{-3}
\end{math}, against the value obtained by Pettini (1999) by assuming 
\begin{math}
     \dot{\rho_{Z}}(z) = \rho_{*}(z)/42     
\end{math}
(see Pettini 1999 and references therein), which is $4.5 \times 10^{6} M_{\odot} Mpc^{-3}$.
The observed contribution from DLA systems and Ly$\alpha$
forest  represents $10\%$ of such amount. This gives rise to the missing metal problem:
where are the metals missing from the cosmic census at high redshift?
Two Possible solutions could be either that metals are hidden in still unseen  Lyman break and SCUBA galaxies
or in warm gas forming (proto-) clusters.
Figure 3b shows the redshift evolution of the comoving density of neutral hydrogen, normalized to the critical density
of the universe. 
Once again, at $z=0$ we normalize our galaxies to the B-band luminosity function.
Our predictions are in reasonable agreement with the observations.
DLA systems are likely to dominate the neutral gas mass density of the high-redshift universe, whereas 
in the local universe, optical surveys indicate that $89\%$ of the neutral hydrogen mass is in spiral galaxies (Rao \& Briggs 1993).
We stress that our calculations depend on the $z=0$ normalization, based on optical observations.
If the $z=0$ estimate of the total HI is much higher than what estimated in the optical bands
(Rosenberg \& Schneider 2002), this could mean that in the local universe the bulk of the 
neutral gas could hide in faint objects below the detection levels of the present-day optical surveys.

\begin{figure} 
\centerline{\includegraphics[width=17.5pc]{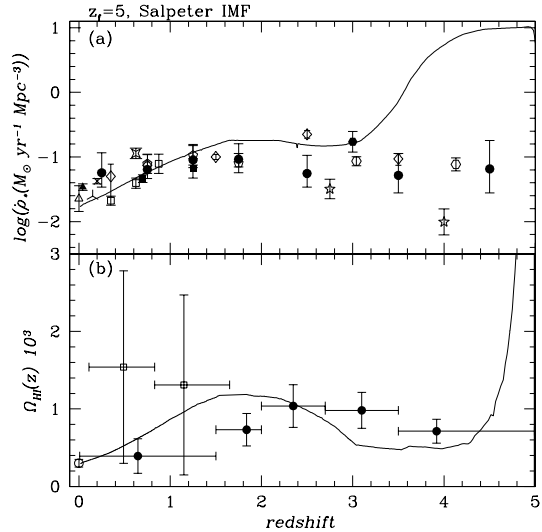}} 
\caption{\tiny
\emph{Upper panel}: Predicted and observed cosmic SFR density in the case of galaxy 
formation occured at $z_{f}=5$ (above). Data taken from Somerville et al. (2001).
\emph{Lower panel}: Predicted and observed
evolution of the neutral H as a function of redshift. 
Open circle: Zwaan et al. (1997); open squares: Rao \& Turnshek
(2000); solid circles: P\'eroux et al. (2001). 
} 
\end{figure}

\end{article}
\end{document}